\documentclass[12pt,aps,superscriptaddress]{revtex4}

\usepackage{graphicx}

\begin{document}

\title{The Quest for the Nuclear Equation of State}

\author{J\"org Aichelin}

\affiliation{SUBATECH, Universit\'e de Nantes, EMN, IN2P3/CNRS, 4 rue
  Alfred Kastler, 44307 Nantes cedex 3, France}

\author{J\"urgen Schaffner-Bielich}

\affiliation{Institut f\"ur Theoretische Physik,
  Ruprecht-Karls-Universit\"at, Philosophenweg 16, 69120 Heidelberg,
  Germany}

\begin{abstract} 
The present status of the efforts to determine the nuclear equation of state
from results of heavy ion reactions and from astrophysical observations is
reviewed.
\end{abstract}

\maketitle

\section{Introduction}

Theory predicts that hadronic matter at finite temperatures and
densities has a rich structure. At moderate temperatures and
densities, below the normal nuclear matter density, $\rho_0$, there
may be a liquid-gas phase transition above which nucleons are not
bound anymore. With increasing temperature nuclear resonances and
mesons appear and nuclear matter becomes hadronic matter. At a
temperature between 165 and 195 $MeV$ lattice gauge calculations
predict (for zero chemical potential) a transition toward a plasma
of quarks and gluons. For finite chemical potentials the transition
temperature becomes even smaller.

Today it is still a challenge to confirm these predictions by
experiments. Even the much simpler question "How much energy is
needed to compress hadronic matter?" has, after 70 years of nuclear
physics, not found a definite answer yet, despite of the importance
of the answer not only for a fundamental understanding of  hadronic
matter but also for the understanding of many astrophysical
observations. This search has been dubbed 'Quest for the hadronic
equation of state' (EoS). Despite of progress in recent years
neither the available data nor the theoretical approaches have
surmounted the difficulties to come to indisputable conclusions for
densities larger than the normal nuclear matter density.

One knows today that there are two possible means to explore the
dependence of the compressional energy density E of hadronic matter
on the density $\rho$ and the temperature T, $E(\rho,T)$: heavy ion
collisions and astrophysical observations. Until 1980 it was even
debated whether in nuclear collisions matter becomes compressed but
the experimental observation of the in-plane flow \cite{gus} and of
the dependence of the  $\pi$ multiplicity on the centrality of the
reaction \cite{sto} showed that nuclei react collectively and that
matter becomes compressed during these reactions. Once the
functional form of $E(\rho,T)$ is known the standard thermodynamical
relations can be employed to study the other thermodynamical
variables like pressure and entropy.

The $E(\rho,T)$ region which can be explored by astrophysical
observations is, however, quite different from that accessible in
heavy ion reactions. The astrophysical objects are usually cold
whereas in heavy ion reactions compression goes along with
excitation. Therefore a detailed knowledge of the entire $E(\rho,T)$
plane and hence of the hadronic interaction is necessary to compare
astrophysical with heavy ion data.

If two nuclei collide with a high energy high densities can be
achieved only for a very short time span ($\approx 10^{-23} s$). Thereafter
the system expands and the density decreases rapidly. The first
exploratory studies of such violent collisions between heavy nuclei
have been carried out at the BEVALAC accelerator at Berkeley/USA, later
the SIS accelerator at GSI/Germany and the AGS accelerator in
Brookhaven/USA have continued and extended this research toward
higher energies. Still higher energies are obtained at the
relativistic heavy ion collider (RHIC) in Brookhaven/USA. Studies of
the properties of hadronic matter at finite chemical potentials will
be possible with the new FAIR project at GSI which
will be operational in 2013.

On the experimental side there are only few observables which give
directly access to the potential between nucleons. The measured
scattering lengths allow to determine the nucleon-nucleon potential in
the different spin and isospin channels at low densities and, via the
so-called $\rho t$ approximation, the binding energy at low
densities. Weizs\"acker has parameterized (above A=40 with a precision
of 1\%) the binding energies of stable nuclei by:
\begin{equation}
E = - a_V A + a_S A^{2/3} + a_CZ^2A^{-1/3}+ a_A \frac{(A-2Z)^2}{A}+
\lambda a_P\frac{1}{A^{3/4}}.
\end{equation}
The first term with  $a_V =15.75 \ MeV$ presents the volume energy,
the second with $a_S= 17.8 \ MeV$ the surface energy, followed by
the Coulomb- and the symmetry energy with $a_c = 0.710 \ MeV$ and
$a_A= 23.7 \ MeV$. The last term is the pairing energy with $a_P =
23.7 \ MeV$ \cite{gre}. $\lambda$ is $-1,0,1$ for odd-odd, odd-even
and even-even nuclei, respectively. From this fit to data we can
conclude that at $\rho_0$ the binding energy per nucleon in nuclear
matter is E/A = -15.75 MeV and hence twice as large as the binding
energy of finite nuclei.

On the theoretical side the difficulty to explore the density
dependence of the compressional energy roots in three facts. First,
it is a many body problem. With increasing density an
increasing number of many-body Feynman diagrams has to be
calculated. Second, the interaction between nucleons is exclusively
phenomenological because it is not yet possible to link it to the
fundamental theory of strong interaction, the Quantum Chromo
Dynamics. The available data allow for different parameterizations
which give, in turn, different density dependencies of the
compressional energy. Third, the analysis of nucleon-nucleon
scattering data reveals that the interaction has a hard core, i.e.
that it becomes infinite or at least very large if the distance
between the nucleons becomes smaller than $a = 0.4 - 0.5 fm$. At
intermediate distances the interaction is moderately attractive.
Potentials based on meson exchange, like the Bonn or the Paris
potential, allow to understand the interaction in terms of different
mesons which are exchanged between the nucleons.

The hard core of the bare nucleon-nucleon interaction makes the
usual concepts of many body physics, like the Hartree Fock mean
field approach, inapplicable because the matrix elements diverge.
The way out of this dilemma are so-called effective interactions,
which are a partial resummation of many-body Feynman diagrams. The
bare interaction is then simply the Born term of such a series.

\begin{figure}
\begin{minipage}[c]{0.4\textwidth}
\includegraphics[width=\textwidth]{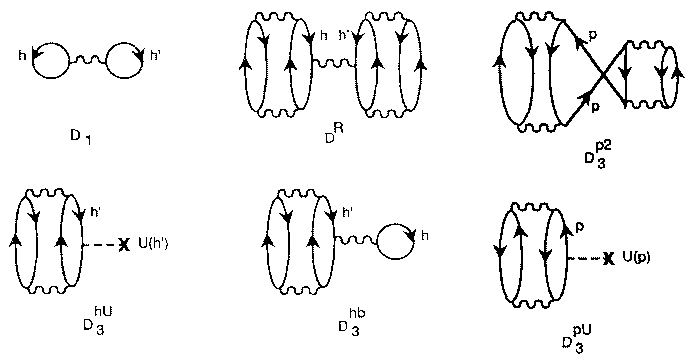}
\end{minipage}
\qquad
\begin{minipage}[c]{0.5\textwidth}
\includegraphics[width=\textwidth]{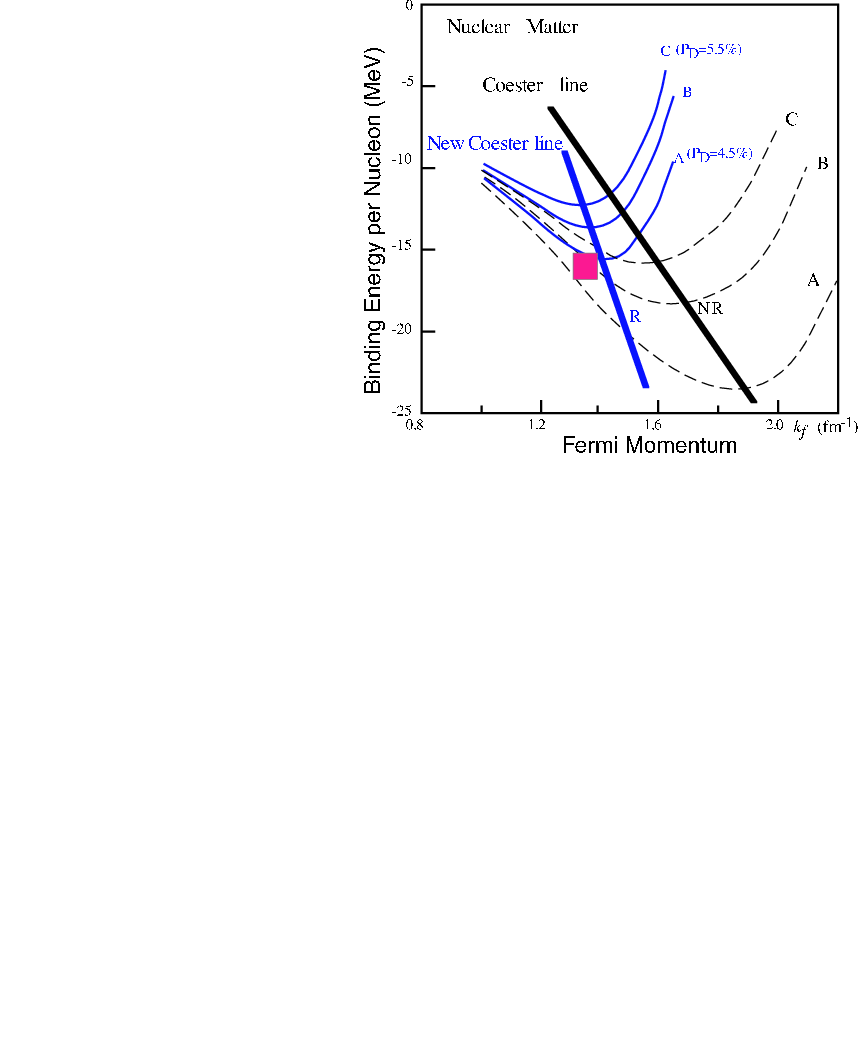}
\end{minipage}
\caption{Left: Different diagrams which contribute to the many-body
calculation of the ground state energy of nuclear matter~\protect\cite{bal90}.
Right:
Binding energy per nucleons as a function of the Fermi momentum in
many-body calculations, after~\cite{bro90}.}
\label{coester}
\end{figure}

The average internuclear distance d, given by
\begin{equation}
\frac{1}{d^3} = \rho = \frac{2k_F^3}{3\pi^2}
\end{equation}
is at $\rho_0$ about three times as
large as the hard core radius a. $k_F$ is the wave number $(k_F=1.42 \
fm^{-1}$) at $\rho_0$. Neglecting the moderate attraction at
intermediate distances each hole line in the many-body Feynman
diagrams contribute a factor $k_F a$ \cite{mat}. To describe nuclear matter
it is therefore appropriate to resum
those many-body diagrams which contain a minimal number of hole
lines. They are presented on the left hand side of fig.~\ref{coester}.

This approach has been developed by Br\"uckner and is called
Br\"uckner G-matrix approach \cite{Bru}. Nonrelativistic
calculations which use different parameterizations of the
nucleon-nucleon potential produce results which all line up along the
so-called Coester line, shown as NR on the right side of fig.
\ref{coester}. Obviously they reproduce neither the experimental
nuclear matter binding energy marked by the rectangle nor the
equilibrium density. Three body (or density dependent two body)
potentials have to be added to bring the calculation in agreement
with data but these additional potentials add also to the
uncertainty of the calculations because their momentum as well as
the density dependence of their strength is not well determined.
Relativistic calculations improve the situation because the inherent
production of virtual nucleon-antinucleon pairs acts like a
repulsive interaction. Also they fall on a common line, marked by R
in fig. \ref{coester}. For the details of the many body approach to
nuclear matter calculations we refer to the excellent review of
Baldo and Maieron \cite{baldo07}.

\begin{figure}
\centerline{
\includegraphics[width=0.8\textwidth]{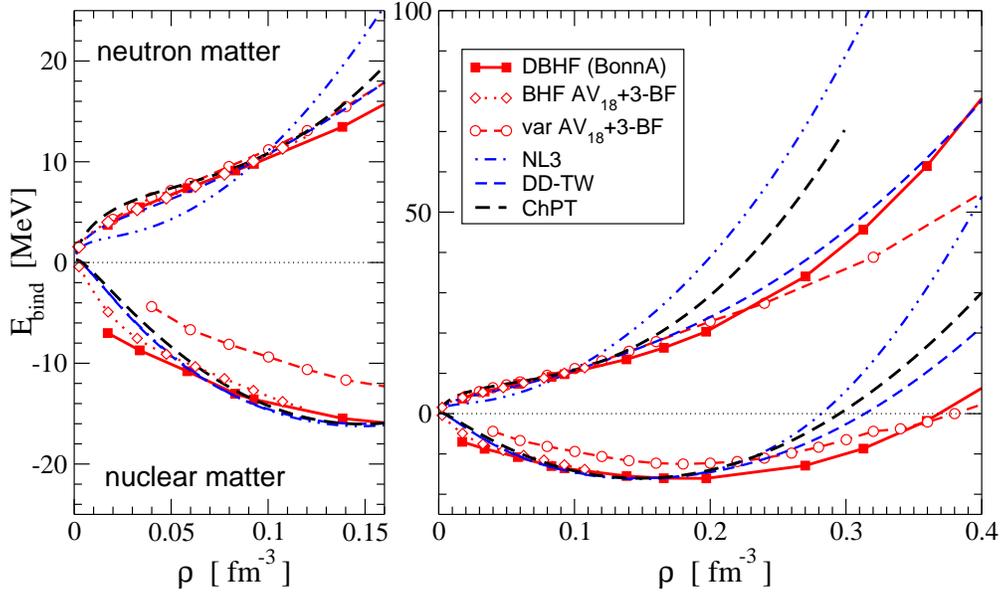}}
\caption{Binding energy per nucleon in nuclear matter and neutron
matter. BHF/DBHF and variational calculations are compared to
phenomenological density functionals (NL3, DD-TW) and ChPT+corr..
The left panel zooms the low density range. The figure is taken from
ref. \cite{WCI} where also the references to the different
calculations can be found. } \label{gmat}
\end{figure}

The state of the art many-body calculations agree relatively well at
$\rho < \rho_0$ being constrained at $\rho\approx 0$ and at
$\rho=\rho_0$, as discussed, but start to deviate substantially for
$\rho > \rho_0 $. This is shown in fig. \ref{gmat}, taken from
\cite{fuc}, which displays the binding energy per nucleon for
neutron matter and nuclear matter predicted by different approaches.
It is evident from the figure that for $\rho > 1.5 \rho_0$ the
present theoretical uncertainty is as large as the binding energy
per nucleon itself.

In order to make progress further experimental information is
needed. Volume oscillations of nuclei \cite{mon,shl,col,vre} would
provide information on the curvature of the binding energy around
the ground state density  and hence allow the determination of the
compressibility modulus
\begin{equation} K = \frac{1}{\kappa} = -V\frac{dp}{dV}= 9
\rho^2 \frac{d^2E(\rho)/A}{(d\rho)^2} |_{\rho=\rho_0}= R^2
\frac{d^2E(\rho)/A}{dR^2}.
\end{equation}
$\kappa$ is the compressibility. Such a
volume oscillation can be induced by the scattering of $\alpha$
particles. Their energy loss $\Delta E$ measures the excitation of
the nucleus which is directly related to K by\cite{dan}
\begin{equation}
\Delta E = \sqrt{\frac{K}{m_N<r^2>_A}}
\end{equation}
where $<r^2>_A$ is the squared radius of the nucleus and $m_N$ the
nucleon mass. A careful analysis of the excitations of different
nuclei shows that the compressibility modulus has bulk, surface,
Coulomb and paring contributions, in analogy with the binding energy
parametrization by Weizs\"acker.

The values found for the volume compressibility in different
non-relativistic and relativistic approaches are around $K = 240$
MeV \cite{mon,shl,col,vre}. Very recently this value has
been questioned because the influence of the surface compressibility
has been underestimated \cite{sha}. This may cause an uncertainty of
30\%.

The volume oscillations are, however, tiny and therefore information
on the compressional energy for $\rho
\gg \rho_0$ cannot be obtained by this method.


\section{The EoS and Heavy-Ion Collisions}


The only way to get on earth to densities well above $\rho_0$ are
high energy heavy ion collisions. The challenge there is to identify
those observables which carry information on the density and on the
compressional energy which is obtained during the reaction and then
to extract robust conclusions. To find these observables is
complicated: The experimental results have to be compared with
theoretical predictions calculated for different assumptions on the
density and compressional energy. A prerequisite for such an
approach are reliable simulation programs which make robust
predictions.

The development of such programs faces a number of problems:
\begin{itemize}
\item[a)] heavy ions are not just a chunk of nuclear matter. Already the
Weizs\"acker mass formula tells us that surface effects are
important.
\item[b)] in heavy ion collisions compression is always accompanied by
excitation which opens new degrees of freedom like resonance
and meson production.
\item[c)] the experimental spectra show that the system does not come even
close to thermal equilibrium during the reaction.
\item[d)] the time evolution of the reaction is strongly influenced by the
production cross section for mesons and nuclear resonances and by
the interaction among these particles. For many reaction channels
the cross sections have not been measured
and for many particles the interaction is not known.
\end{itemize}
In the last decades transport theories have been developed which
cope with these challenges. Based on quantum molecular dynamics
\cite{iqmd,urqmd,qmd} or the quantum version of the Boltzmann
equation\cite{hsd,rbuu} these approaches simulate heavy ion
reactions from the beginning, when projectile and target are still
separated, to the end, when the particles are registered by the
detectors. In the molecular dynamics approaches
\cite{iqmd,urqmd,qmd} nucleons  are represented by Gaussian wave
functions with a constant, time independent width. Hence the
Wigner density of a nucleons reads as
\begin{equation}
 f_i (\vec{r},\vec{p},t) = \frac{1}{\pi^3 \hbar^3 }
 {\rm e}^{-(\vec{r} - \vec{r}_{i} (t) )^2  \frac{2}{L} }
 {\rm e}^{-(\vec{p} - \vec{p}_{i} (t) )^2  \frac{L}{2\hbar^2}.  }
\label{fdefinition}
\end{equation} The total one particle Wigner
density is the sum of the Wigner densities of all nucleons. The
particles move according to Hamilton's equations of motion
\begin{equation}
\dot{r_i}=\frac{\partial <H>}{\partial p_i} \qquad
\dot{p_i}=-\frac{\partial <H>}{\partial r_i}.
\end{equation}
The expectation value of the total Hamiltonian in this approach is approximated
by
\begin{eqnarray}
\langle H \rangle &=& \langle T \rangle + \langle V \rangle
\nonumber \\ \label{hamiltdef} &=& \sum_i \frac{p_i^2}{2m_i} +
\sum_{i} \sum_{j>i}
 \int f_i(\vec{r},\vec{p},t) \,
V^{ij}  f_j(\vec{r}\,',\vec{p}\,',t)\, d\vec{r}\, d\vec{r}\,'
d\vec{p}\, d\vec{p}\,' \quad.
\end{eqnarray}
The baryon-potential consists of the real part of the $G$-Matrix
which is supplemented by the Coulomb interaction between the charged
particles. The former can be further subdivided in a part containing
the contact Skyrme-type interaction only, a contribution due to a
finite range Yukawa-potential, a momentum dependent part and a
symmetry energy term depending of the difference between proton and neutron
densities \cite{iqmd}. In infinite
matter the interaction is reduced to
\begin{equation}
E/N = m_N + E_{kin} +
\alpha\frac{\rho}{\rho_0} + \beta(\frac{\rho}{\rho_0})^\gamma+
S\frac{\rho}{\rho_0}[\frac{\rho_n-\rho_p}{\rho}]^2 +\epsilon f(p).
\label{eos}
\end{equation}
The imaginary part of $G$-Matrix acts as an elastic cross section
which is complemented by inelastic cross sections.

This parametrization of the potential uses the minimal number of
parameters because in infinite matter two of the three parameters
$\alpha, \beta, \gamma$ are determined by the requirement that the
binding energy is minimal at $\rho_0$ and equal to -15.75 $MeV$. The
third parameter can be expressed in terms of the compressibility
modulus K. This restrained ansatz is necessary as long as no
observables have been identified which allow to fix further
parameters like for example the (not necessarily linear) density
dependence of the symmetry energy. $\epsilon f(p)$ is determined
from optical potential measurements in p-A reactions.

For heavy ion collisions the situation is even more complicated
because there is no constant density in the reaction zone and
therefore the density has to be calculated by summing up the Wigner
densities. The local density  is therefore dependent on the width of
the (Gaussian) wave functions, a parameter which is only vaguely
controlled by nuclear surface properties.

In order to characterize the results the following strategy has been
employed: The parameters of the two and three body interactions
which are actually employed in the calculations, are fixed by the
requirement that they agree in infinite matter with a given set of
$\alpha,\beta,\gamma$. This allows to characterize the potential
parameters by a compressibility modulus: 'Soft' means K=250 $MeV$
and 'Hard' means K= 380 $MeV $.

\begin{figure}
\centerline{\includegraphics[width=0.8\textwidth]{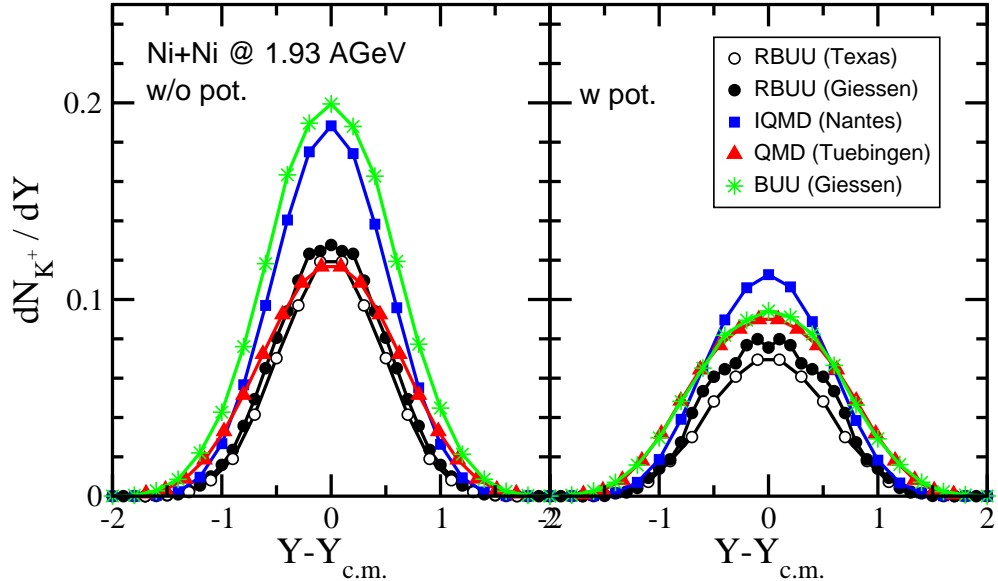}}
\caption{$K^+$ rapidity distributions in central (b=1 fm) Ni+Ni
reactions at 1.93 AGeV from various transport models in their
default versions: RBUU (Texas, open circles), RBUU (Giessen, full
circles), IQMD (Nantes, full squares), QMD (T\"ubingen, full
triangles) and BUU (Giessen). The left figure shows results without
kaon in-medium potentials  while the right one includes  potentials.
The figures are taken from refs. \cite{trento03,fuchs}.}
\label{dndy_2}
\end{figure}

In the last two decades simulations using these transport theories
became  an indispensable tool to interpret the results of heavy ion
collisions from $E_{kin} > 50 \ MeV$ up to the highest beam
energies. Because they predict the entirety of the experimental
observables on an event by event basis, correlations can be
identified and cross checks can be easily performed. These
simulations have identified two observables which are sensitive to
the compressional energy at $\rho \gg \rho_0$:
\begin{itemize}
\item[a)] the in-plane flow of nuclei in semi-central heavy ion
  collisions
\item[b)] the production of $K^+$ mesons at subthreshold and threshold
  energies.
\end{itemize}
For semi-central collisions already hydrodynamical calculations have
predicted an in-plane flow. The transport theories allowed for a
quantitative prediction of this collective phenomenon. For a review
we refer to \cite{st86}. When two nuclei collide the interaction
zone has a higher density than the surrounding spectator matter. The
density gradient creates an energy gradient and hence a force, F{\bf
e} (see eq. \ref{eos}).  ${\bf e}$ lies (almost) in the reaction
plane and is (almost) perpendicular to the beam direction:  ${\bf
e}=e_x$ . The force changes the momentum of the nucleons at the
interface between participant and spectator zone by
\begin{equation}
\Delta p_x \approx \frac{dV(x)}{dx} r_0
A^{1/3} \frac{m_N}{<p_z^{cm}>} \label{pxdir}
\end{equation}
where $<p_z^{cm}>$ is the momentum of a nucleon in beam direction in
the center of mass system and $ r_0 A^{1/3}$ the length of the
nucleus. The nucleons at the surface of the interaction zone, where
$|\frac{dV(x)}{dx}|$ is largest, transfer this momentum to the
nucleons which are around creating a collective in-plane flow. For
projectile and target nucleons the momentum transfer is of opposite
direction and can be measured because projectile (target) nucleons
end up at forward (backward) rapidity. It is evident from
eq.~(\ref{pxdir}) that $\Delta p_x$ depends via $\frac{dV(x)}{dx}$
on the compressional energy and the simulation programs have
verified this dependence. The difficulty is that $<\Delta p_x>$ is
tiny and the difference of $< \Delta p_x>$ for different
parameterizations of the compressional energy is even smaller. A
quantitative prediction depends crucially on the ability of the
simulations programs to simulate very accurately not only the bulk
but also the properties of the surface where $\frac{dV(x)}{dx}$ is
largest. Presently the tiny difference of $\frac{dV(x)}{dx}$ for two
different equations of state in one program is smaller than the
difference of $\frac{dV(x)}{dx}$ between two different programs
which use the same EoS and which predict the same bulk properties.
Therefore it is premature to make quantitative predictions of the
compressional energy based on the observed in-plane flow \cite{ant}.

For the second method to study compressional energies at high
densities, on the contrary, the results of different simulation
programs have converged \cite{trento03,fuchs}. This is shown in fig.
\ref{dndy_2} where the $K^+$ rapidity distributions for central (b=1
fm) Ni+Ni reactions at 1.93 $AGeV$ from various transport models are
displayed. The different results in their default versions come from
different assumption on only vaguely known input quantities: The
N$\Delta$ cross section, the $\Delta$ lifetime in matter and the
strength of the KN potential. Once the same input is used the
results agree quite well.

\begin{figure}
\centerline{\includegraphics[width=0.6\textwidth]{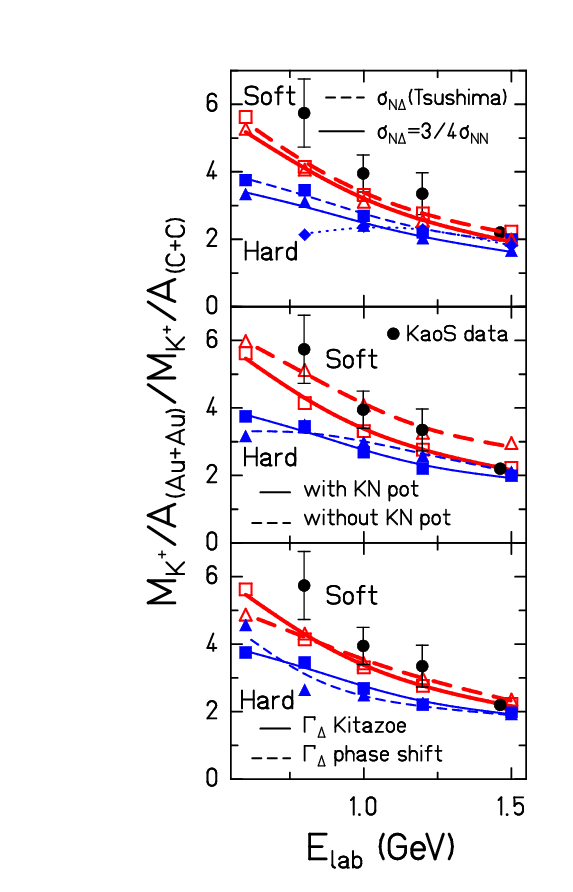}}
\caption{Comparison of the measured excitation
function of the ratio of the $K^+$ multiplicities per nucleon
obtained in Au+Au and in C+C reactions (Ref.~\cite{sturm}) with
various calculations. Results of simulations with a hard EoS are
shown as thin (blue) lines, those with a soft EoS by thick (red) lines.
The calculated values are given by symbols, the lines are drawn to
guide the
eye. On top, two different versions of the $N\Delta \rightarrow
K^+\Lambda N$ cross sections are used. One is based on isospin
arguments \cite{ko}, the other is determined by a relativistic tree
level calculation \cite{tsu}.
Middle: IQMD calculations with and without $KN$ potential are
compared. Bottom: The influence of different options for the life
time of  $\Delta$ in matter is demonstrated. The figure is taken
from ref.\cite{Hart_eos}.} \label{kratio}
\end{figure}

$K^+$ mesons produced far below the $NN$ threshold cannot be created
in first-chance collisions between projectile and target nucleons.
They do not provide sufficient energy even if one includes the Fermi
motion. The necessary energy for the production of a $K^+$ meson in
the $NN$ center of mass system is 671 $MeV$ because in addition to
the production of a kaon a nucleon has to be converted into a
$\Lambda$ to conserve strangeness. Before nucleons can create a
$K^+$ at these subthreshold energies, they have to accumulate
energy. The most effective way to do this is to convert a nucleon
into a $\Delta$ and to produce in a subsequent collision a $K^+$
meson via $\Delta N \rightarrow N K^+ \Lambda$. Two effects link the
yield of produced $K^+$ with the density reached in the collision
and the stiffness of the compressional energy. If less energy is
needed to compress matter (i) more energy is available for the $K^+$
production and (ii) the density which can be reached in these
reactions will be higher. Higher density means a smaller mean free
path and therefore the time between collisions becomes shorter. Thus
the $\Delta$ has an increased chance to produce a $K^+$ before it
decays. Consequently, the $K^+$ yield depends on the compressional
energy. At beam energies around 1 $AGeV$ matter becomes highly
excited and mesons are formed. Therefore this process tests highly
excited hadronic matter. At beam energies $> 2~AGeV$ first-chance
collisions dominate and this sensitivity is lost. The simulations
verify that  the $K^+$ in Au+Au collisions at 1.5 $AGeV$ are indeed
produced at around $2\rho_0$ \cite{Hart_eos}.

\begin{figure}
\centerline{\includegraphics[width=0.6\textwidth]{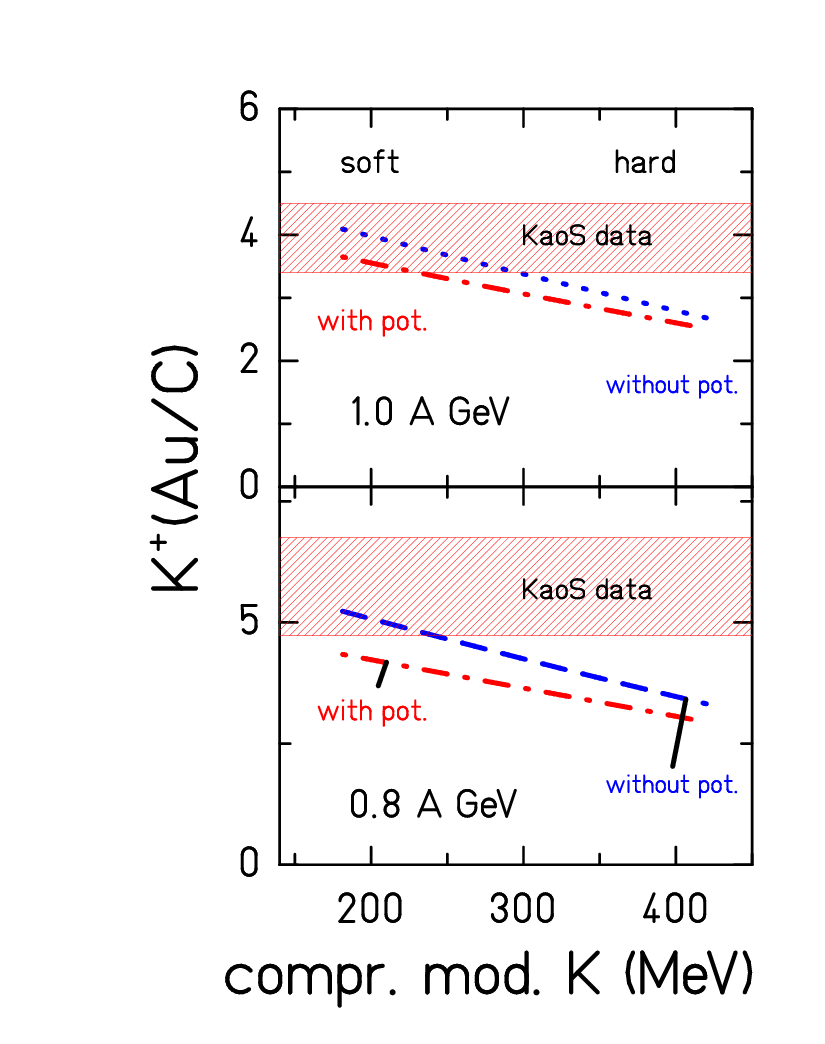}}
\caption{The double ratio
  $[M/A({\rm \mbox{Au+Au}})] / [M/A({\rm \mbox{C+C}})]$
calculated within the IQMD model (with and without KN potential) as
a function of $K$ for two beam energies, 0.8 (top) and 1.0 $AGeV$
(bottom). The experimental values are given as a band and allow to
estimate upper limits for the compressibility modulus $K$ as
described in the text.} \label{IQMD_ratio}
\end{figure}

The $K^+$ mesons behave as a quasi particle even at high densities
\cite{kor} and can therefore be propagated as the baryons in these
simulations programs. The largest uncertainty of the $K^+$
production in heavy ion collisions is the only theoretically
calculated $\Delta N$ production cross section \cite{ko,tsu} which
is dominating. Its influence on the observables can be minimized by
analyzing ratio of the multiplicity in light and heavy symmetric
systems. It is further constrained by the excitation function of the
$K^+$ multiplicity. This ratio is displayed in fig. \ref{kratio}.
All three graphs show calculations with a soft and a hard EoS. For
the three calculations little known or unknown input parameters are
varied ($N\Delta$ cross section, top, KN potential , middle, and
$\Delta$ life time, bottom) to see whether the conclusion is robust.
We see that different assumptions on these input quantities do not
invalidate the conclusion that the data are incompatible with the
assumption that the EoS is soft.

\begin{figure}
\centerline{\includegraphics[width=0.6\textwidth]{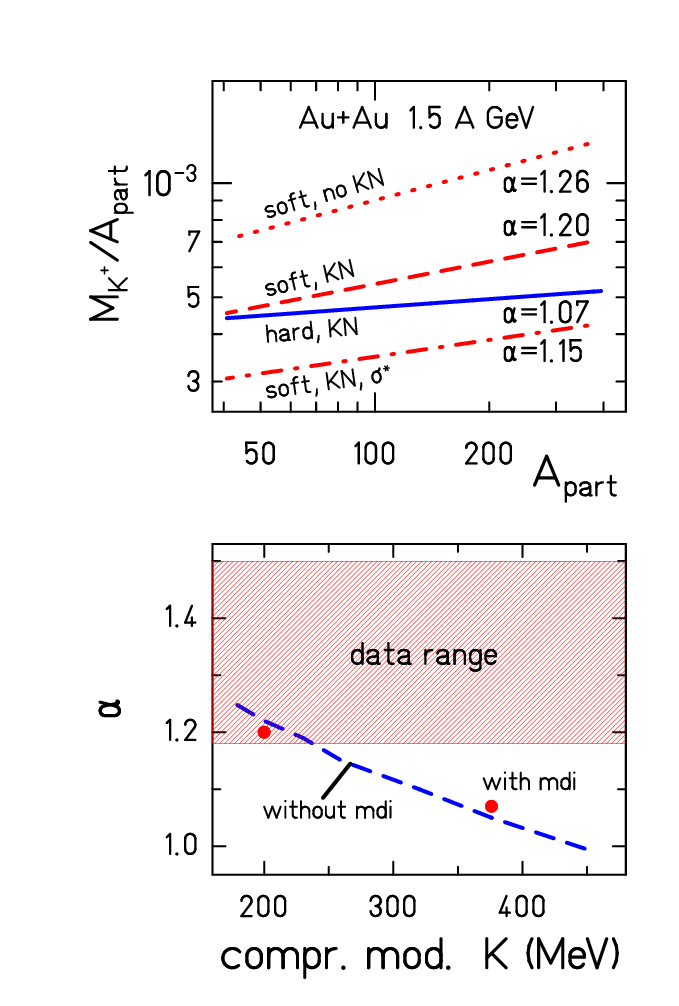}}
\caption{Dependence of the $K^+$ scaling on the EoS. We present this
dependence in form of $M_{K^+}=A_{{\rm part}}^\alpha$. On the top
the dependence of $M_{K^+}/A_{{\rm part}}$ as a function of $
A_{{\rm part}}$ is shown for different options: a hard EoS with $KN$
potential (solid line), the other three lines show the result for a
soft EoS, without $KN$ potential and $\sigma(N\Delta)$ from
Tsushima~\cite{tsu} (dotted line), with $KN$ potential and the same
parametrization of the cross section (dashed line) and with $KN$
potential and $\sigma(N\Delta) = 3/4 \sigma(NN)$. On the bottom the
fit exponent $\alpha$ is shown as a function of the compressibility
modulus for calculations with momentum-dependent interactions (mdi)
and for static interactions ($\epsilon$=0, dashed line).}
\label{partff}
\end{figure}

Fig. \ref{IQMD_ratio} shows the double ratio as a function of the
compressibility modulus K for two beam energies with and without KN
potential. Also for this observable data are not compatible with a
hard EoS.

This conclusion is confirmed by another, independent observable, the
centrality dependence of the $K^+$ production. It can be described
by $M_{K^+} \propto A_{{\rm part}}^\alpha$. The value of $\alpha$
depends on the choice of the input parameter, as shown in fig.
\ref{partff}, top. Again, only a soft EoS can describe the data, as
can be seen in the bottom part, where we display the result with and
without a momentum dependent interaction ($\epsilon = 0$ in eq.
\ref{eos}).

Thus heavy ion reactions, in which densities up to $3\rho_0$ are
obtained and in which almost all $K^+$ are produced at densities
well above $\rho_0$, the experimental results are only compatible
with a soft EoS. The value of K obtained from high energy heavy ion
collisions agrees well with that extracted from the analysis of
density vibrations around $\rho_0$. This has to be considered as
accident because the density dependence of the compressional energy
can be more complicated than suggested by the simple parametrization
of eq. \ref{eos}. The precision of present day experiments and
theory does not allow for a determination of more than one parameter
which has been traditionally expressed as the compressibility
modulus K at $\rho_0$. One has to keep in mind that in heavy ion
reactions this value of K has been extracted in a complicated way
from a very excited non equilibrium system where mesons and nuclear
resonances are present. Conclusions on the compressibility modulus
of nuclear matter at large densities and a small temperature have
therefore to be drawn with great caution. Nevertheless, this
analysis presents presently the only robust information on the
compressional energy of hadronic matter well above normal nuclear
matter density which has been obtained from heavy ion experiments.


\section{The EoS and Astrophysics}


The nuclear equation of state as determined from heavy-ion experiments
has crucial impacts in high-density astrophysics in particular on the
physics of neutron stars, core-collapse supernovae, and neutron star
mergers.

Neutron stars are the final endpoint of stellar evolution of stars
more massive than about eight solar masses. Those stars end in a
spectacular core-collapse supernova which outshines for a brief moment
of time even the light of an entire galaxy. Matter is compressed to
extreme densities, energy densities above normal nuclear matter
density are reached in the collapse of the degenerate core of the
supermassive progenitor star. Only the strong repulsive interactions
between nucleons can prevent the further collapse to a black hole
providing the enormous pressure necessary to withstand the pull of
gravity. A stable proto-neutron star is formed with initial
temperatures of about 20 to 50 MeV.  It is important to realize that
the degeneracy pressure alone is not enough to ensure the stability of
the proto-neutron star, it is essential that the nuclear equation of
state controls the bounce back of material during the collapse of
matter. A shock wave is built up due to the strong repulsion of
nucleons which is moving outwards. The stable hot proto-neutron star cools
down within about one minute by the emission of neutrinos from its
surface which have been traveling by a random walk from the core. The
temperature drops down to less than 1 MeV within that first
minute. The temperature is now so low in comparison to the Fermi
energy of the nucleons in the core region, that temperature effects
can be safely ignored afterwords. A cold neutron star is born which is
still emitting neutrinos in a wind.

\subsection{The nuclear EoS, supernovae and neutron star mergers}

The conventional mechanism to be believed to be responsible for a
successful explosion is the so called neutrino driven explosion
\cite{Bethe:1984ux} which happens on a timescale of less than a second
after the bounce. The shock wave generated by the bounce of matter is
not energetically enough to plow through all the material of the
progenitor star and stalls at a few hundred kilometers. The neutrinos
emitted from the hot proto-neutron star carry an enormous energy, in
total $10^{53}$ erg, which is about two orders of magnitudes larger
than the energy contained in the outflowing material of observed
supernovae. Therefore, if just a small fraction of the energy of
neutrinos is transferred to the stalled shock front, it could be
revived so that a successful explosion occurs.

However, until recently, no successful explosion could be achieved
even with improved models, in particular with respect to the treatment
of the propagation of neutrinos.  One dimensional supernova
simulations were inherently unsuccessful in achieving sufficient
explosion energies so far.  New mechanisms have been suggested for a
successful supernova explosion as the standing accretion shock
instability (SASI) \cite{Janka:2004an,Buras:2005tb} or acoustic
oscillations from the proto-neutron star \cite{Burrows:2005dv}, which
were observed when treating the dynamics of supernova evolution in
unconstrained multi-dimensional simulations. The results demonstrate
that nonradial hydrodynamic instabilities, which can help to support
explosions, depend on the underlying nuclear equation of state
\cite{Janka:2004an}. In particular, the stiffness of the nuclear
equation of state affects the time variability on the neutrino and
gravitational wave signal with larger amplitudes as well as higher
frequencies for more compact newly born neutron stars. Information on
the nuclear equation of state is therefore inherently imprinted on the
neutrino and gravitational wave emission \cite{Marek:2008qi}. For
simulations taking into account rotation, the effects of the nuclear
equation of state on these astrophysical observables seems to be much
smaller though \cite{Dimmelmeier:2008iq}.

The present situation is described poignantly by the final statement
in the abstract of the recent review on the theoretical status of
core-collapse supernovae in ref.~\cite{Janka:2006fh}: `The explosion
mechanism of more massive progenitors is still a puzzle. It might
involve effects of three-dimensional hydrodynamics or might point to
the relevance of rapid rotation and magnetohydrodynamics, or to still
incompletely explored properties of neutrinos and the high-density
equation of state.'

Also for neutron star mergers as well as collisions of neutron stars
with black holes impacts of the nuclear equation of state have been
observed in numerical simulations. In
\cite{Rosswog:1999,Rosswog:2000}, it was found that the amount of
material loss to the interstellar medium for merging neutron stars
depends strongly on the stiffness of the nuclear equation of
state with corresponding implications for element synthesis in the
r-process.  The peak in the gravitational wave spectrum is highly
sensitive to the nuclear equation of state and on the total mass of
the binary system. The total mass of the binary neutron star system
could be determined from the inspiral chirp signal so that the
frequency of the postmerger signal serves a sensitive indicator of the
properties of the high-density nuclear equation of state
\cite{Oechslin:2006uk,Oechslin:2007gn}.
For neutron stars being swallowed by black holes the
complete dynamics is governed by the mass loss and how the neutron
star reacts to it. A stiff equation of state causes an episodic mass
transfer over many orbits which is visible in the gravitational wave
signal. On the contrary, for a soft polytropic equation of state it was
observed that the neutron star was ripped apart at the first encounter
with the black hole \cite{Rosswog:2004,Rosswog:2007}.

All of the above mentioned newer investigations have been performed
with basically two different nuclear equation of state which are
constructed in such a way that they are suitable for astrophysical
applications: the one of Lattimer and Swesty \cite{Lattimer:1991nc}
with a Skyrme-type interaction and the one by Shen et al.\
\cite{Shen98} using a relativistic mean-field theoretical model. Both
of them are purely nucleonic in nature and at present the only ones
available in the modern literature. Astrophysical studies of the
effects from quark matter have been hampered by this fact and only a
few exploratory investigations have been performed. In ref.\
\cite{Oechslin:2004yj} neutron star mergers have been calculated with
the use of a simple quark matter equation of state. Effects from the
presence of quark matter have been seen in the collapse behavior of
the merger remnant and in the gravitational wave signal. In ref.\
\cite{Gentile:1993ma} the formation of quark matter produced a second
shock wave but the calculation was performed without any neutrino
transport.  The failed explosion and the collapse of heavy progenitor
stars to a black hole have been studied in ref.\
\cite{Yasutake:2007st,Nakazato:2008su} with effects of a phase
transition to quark matter. Quark matter appeared at quite high
densities so that that collapse to a black hole could not be stopped
and happened even faster than for the case without quark matter. No
second shock wave has been seen in these simulations.

Recently, it was demonstrated in a detailed one-dimensional
computation with full treatment of the neutrino transport that the
formation of a quark-gluon plasma shortly after bounce produces an
accretion shock at the surface of the quark matter core
\cite{Sagert:2008ka}. The second shock front travels outwards and is
so energetic that it passes over the stalled first shock and achieves
a successful explosion with quite large explosion energies. The
presence of the second shock can be observed from the temporal profile
of the neutrino emission from the supernova. The first shock from the
bounce of nuclear matter produces a peak of neutrinos as the matter is
neutronized. A second neutrino burst is generated by the emission of
antineutrinos when the second shock formed from the quark matter core
runs over hadronic matter and neutrons are transformed back to
protons. This second peak in the neutrino spectra serves as a signal
for the presence of a strong phase transition, the time delay compared
to the first neutrino peak and the height of the peak gives insights
onto the location and the strength of the phase transition line in the
QCD phase diagram.

\subsection{The nuclear EoS and compact stars}

The mass-radius curve of neutron stars formed in core-collapse
supernovae is entirely determined by the nuclear equation of state,
with only minor corrections from rotation.  The basic structure
equations in spherical symmetry for compact stars are given by solving
the equations of General Relativity for a static and spherically
symmetric metric and result in the Tolman-Oppenheimer-Volkoff (TOV)
equations \cite{Tolman34,Tolman39,OV39}
$$
\frac{dP}{dr} = - G \frac{M_r \epsilon}{r^2}
\left(1 + \frac{P}{\epsilon}\right)
\left(1 + \frac{4 \pi r^3 P}{M_r}\right)
\left(1 - \frac{2 G M_r}{r}\right)^{-1}
$$
with the mass conservation equation
$$
\frac{dM}{dr} = 4 \pi r^2 \epsilon
$$
There are three relativistic correction factors compared to the
Newtonian expression for the hydrodynamic structure equations for
ordinary stars. One corrections factor is present for the mass $M_r$
contained within the radius $r$ and one for the energy density, where
effects from the pressure are taken into account. The third correction
factor originates from the outside solution and modifies the radius
with the Schwarzschild factor, in particular close to the
Schwarzschild radius $R_s=2GM$ where $M$ is the total gravitational
mass of the star.

The maximum mass of white dwarfs is controlled by the Fermi pressure
of electrons. It is well known that the maximum mass of a neutron star
must be determined by the repulsive nature of the nuclear force, not
by the Fermi pressure of nucleons. The maximum mass of a neutron star
just supported by Fermi degeneracy pressure of a free gas of neutrons
results in a maximum mass of just $0.7M_\odot$ \cite{OV39}.  The mass
of the Hulse-Taylor pulsar has been measured to be $(1.4414\pm 0.0002)
M_\odot$ \cite{Weisberg:2004hi}, which is more than a factor two
larger. Such a large neutron star mass can only be explained by
including effects from strong interactions. One can adopt an inversion
procedure, as first outlined by Gerlach \cite{Gerlach68}, which
relates the mass-radius curve of neutron stars to the underlying
nuclear equation of state \cite{Lindblom92}. Hence, the knowledge of
the masses and radii of compact stars gives immediately a unique
constraint on the properties of dense nuclear matter.

In the near future, one could not only derive properties of matter
under extreme conditions by spectroscopy or by the detection of
neutrinos but also by 'listening' to astrophysical events by the
detection of gravitational waves. If there exists a rigid phase in the
core of neutron stars, pulsars could pertain a slight deformation so
that they wobble thereby emitting characteristic gravitational waves.
More spectacular signals are expected from the merging of two neutron
stars. Several double neutron star systems have been discovered in our
galaxy, the best known is the Hulse-Taylor binary pulsar PSR 1916+13.
Gravity is so strong in those double neutron star systems, that a
significant amount of energy is lost by the emission of gravitational
waves. The two neutron stars spiral inwards and are prone to collide
with each other in a spectacular astrophysical event. The
gravitational wave signal consists basically of three parts. First a
rising chirp occurs just before the two neutron stars merge which is
determined by the compactness of the neutron stars, the ratio of the
mass to radius. The middle part is controlled by the nuclear equation
of state where the actual merger of the two neutron stars is
happening. Finally, the system collapses to a deformed black hole
which is ringing down by the emission of gravitational waves. The
middle part of the gravitational wave pattern is the least well known
due to our limited knowledge of the high-density nuclear equation of
state. However, this fact serves as an opportunity to learn more about
it by measuring gravitational waves. Today several gravitational wave
detectors are in operation, as LIGO, VIRGO, TAMA, GEO600. The LIGO
collaboration just published new limits on the emission of
gravitational waves from the observation of pulsars which are now at
the so-called spin-down limit \cite{Abbott:2007ce}. Pulsars can loose
their rotational energy and spin-down by either the emission of
electromagnetic radiation or by gravitational waves. LIGO is now at
the limit to measure gravitational waves of the energy scale given by
the loss of rotational energy of the crab pulsar. In just a few years
LIGO will enhance its sensitivity by a few factors with the LIGO+
upgrade and advanced LIGO, so that the possible emission of
gravitational waves from single wobbling pulsars can be tested.

\begin{figure}
\centering
\includegraphics[width=0.55\textwidth]{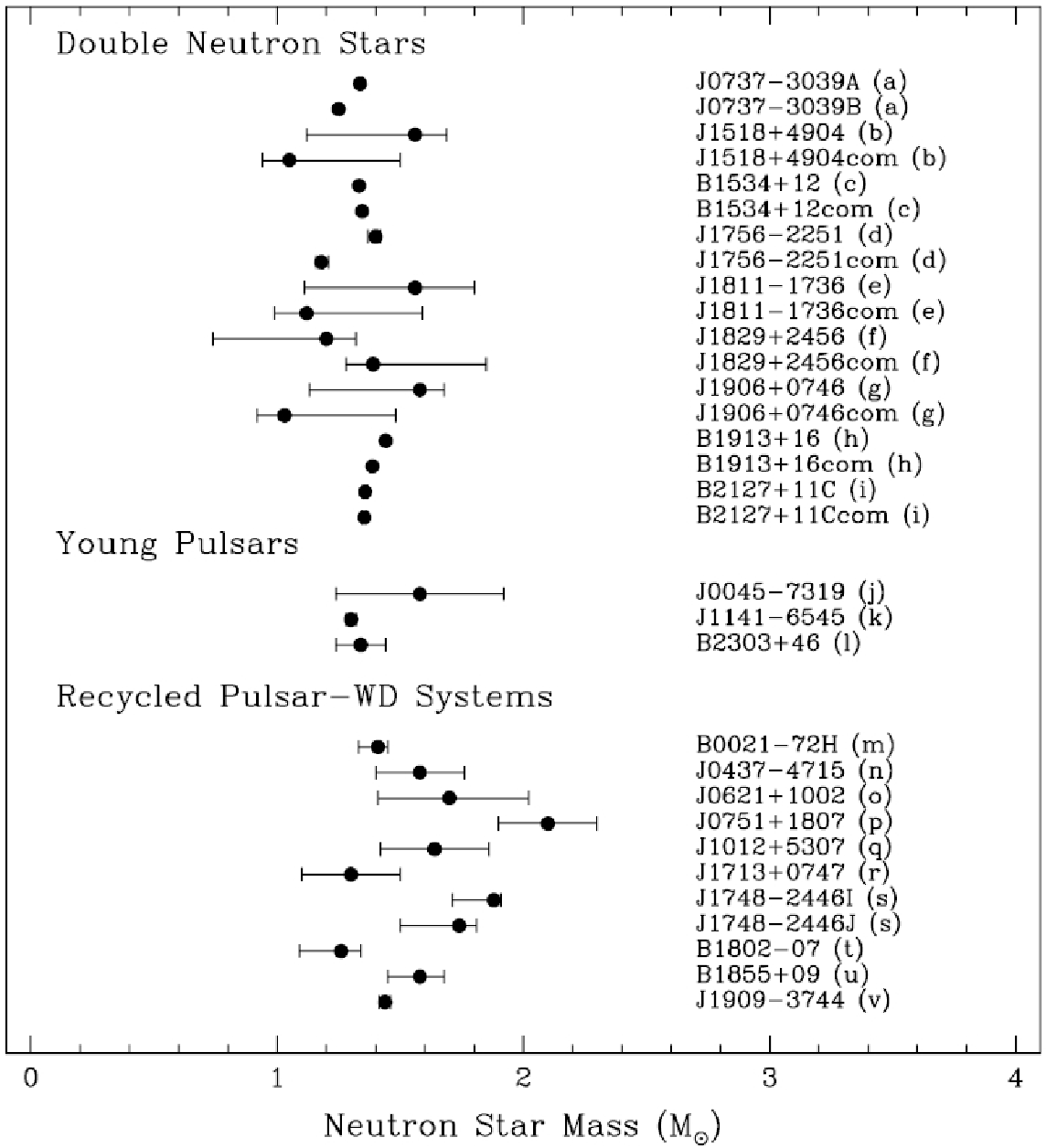}
\caption{Mass measurements from pulsars, rotation-powered neutron
  stars (taken from \cite{Stairs:2006yr}). Note that the mass of the pulsar
  PSR J0751+1807 has been corrected from $M=(2.1\pm0.1)M_\odot$
  downwards to $(1.26\pm 0.14)M_\odot$, see \cite{Nice:2008}.}
\label{fig:masses}
\end{figure}

There are more than 1700 pulsars, rotation-powered neutron stars,
discovered within our local group of galaxies as listed in the ATNF
pulsar database \cite{Manchester:2004bp}. Several binary systems are
known, where a neutron star has a companion, be it a main-sequence
star, a white dwarf or another neutron star. For one system, the
double pulsar J0737-3039A/B, the radio pulses of both neutron stars
have been detected \cite{Lyne:2004cj}.  The masses of pulsars are
determined by measuring corrections from General Relativity for the
orbital parameters as deduced from the timing of the pulsar's radio
signal.  Compilations of neutron star masses have found a rather
narrow range of $1.35\pm 0.05$ \cite{Thorsett99}.
Fig.~\ref{fig:masses} shows a recent compilation of pulsar mass
measurements from \cite{Stairs:2006yr}. The smallest neutron star mass
measured so far is $M = (1.18\pm 0.02) M_\odot$ for the pulsar J1756-2251
\cite{Faulkner:2005}, the heaviest most reliably determined at present
is the one of the Hulse-Taylor pulsar with $(1.4414\pm 0.0002)
M_\odot$ \cite{Weisberg:2004hi}.

In recent pulsar observations \cite{Stairs:2006yr} much larger values
than have been reported but have to be taken with caution. The mass of
the pulsar J0751+1807 was corrected from $M=(2.1\pm 0.1)M_\odot$
\cite{Nice:2005fi} to $(1.26\pm 0.14)M_\odot$ as new data became
available \cite{Nice:2008}. The mass of the neutron star in Vela X-1
has been extracted to be not less than $(1.88\pm 0.13)M_\odot$ for an
inclination angle of 90 degrees, which relies on the radial velocity
measurement of the optical companion star with possible systematic
errors \cite{Quaintrell03}. There are a series of measurements of
extremely massive pulsars in globular clusters, where just the
periastron advance has been determined but not the inclination angle
of the orbit \cite{Ransom:2005ae,Freire:2007jd,Freire:2007xg}. For the
pulsar PSR J1748-2021B a mass of $(2.74\pm 0.21)M_\odot$ is reported
by using a statistical analysis for the inclination angle
\cite{Freire:2007jd}. We stress that this is not a direct mass
measurement and that a second relativistic correction from General
Relativity has to be determined before one can draw any conclusion
about the true pulsar mass.  There is a recent mass measurement by a
highly excentric pulsar, PSR J1903+0327, with $(1.74\pm 0.04)M_\odot$
\cite{Champion:2008}, but over only a time period of 1.5 years so that
possible effects from proper motions affecting the mass measurement
can not be excluded. The other high mass measurement reported for the
pulsar J0437-4715 of $M=(1.76\pm 0.20)M_\odot$ \cite{Verbiest:2008gy}
has a lower $2\sigma$ bound than the one of the Hulse-Taylor pulsar.

\begin{figure}
\centering
\includegraphics[width=0.7\textwidth]{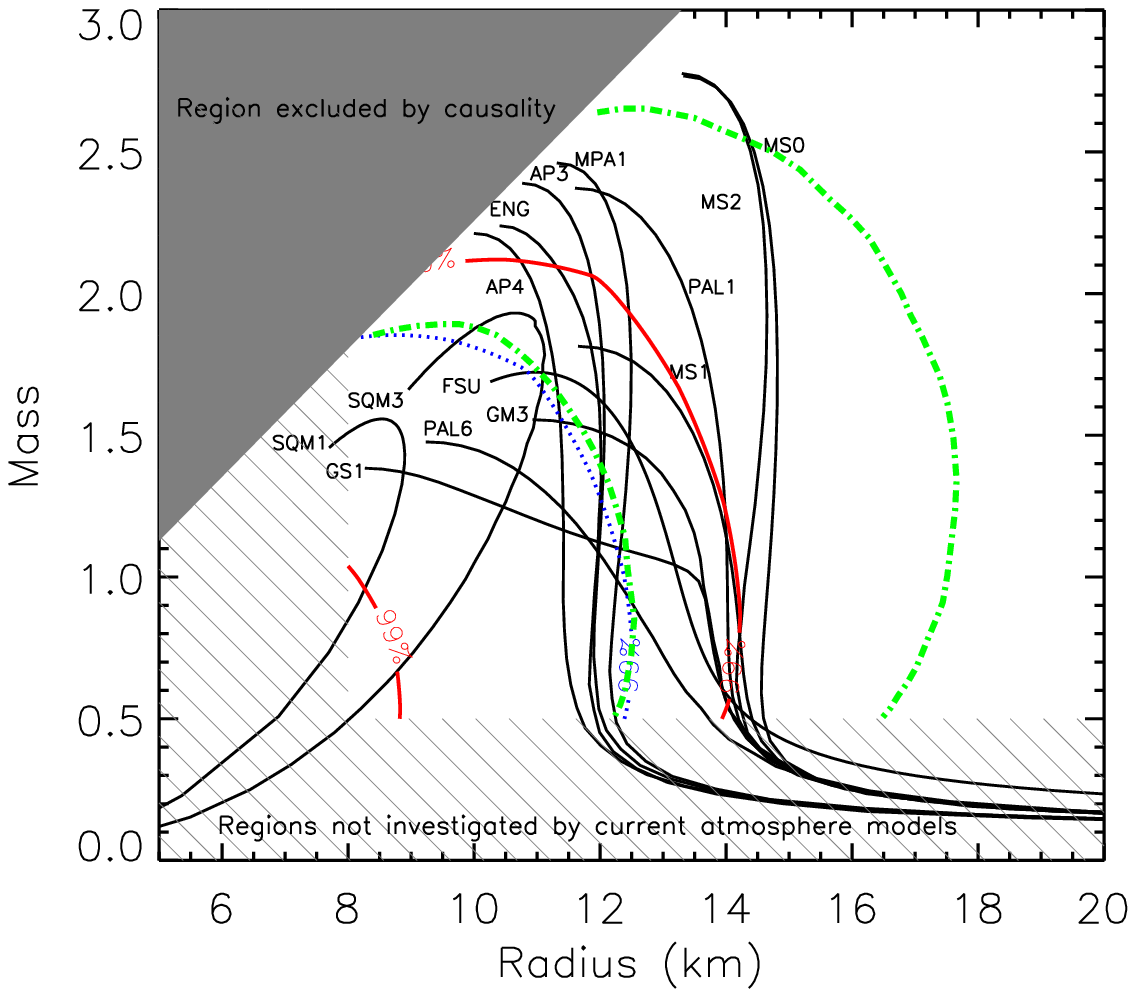}
\caption{Mass-radius constraints from spectral modeling of pulsars in
  globular clusters (from \cite{Webb:2007tc}). The mass-radius curve
  has to pass through the region bounded by the two lines. Three
  different mass-radius constraint have been shown for neutron stars
  in the globular cluster M13 (dotted line), in $\omega$~Cen (solid
  lines) and for X7 in 47 Tuc (dash-dotted lines). The various
  mass-radius curves for different nuclear equation of states shown
  are taken from \cite{Lattimer:2006xb}. The curves starting from the
  origin are for pure quark stars without an hadronic mantle which are
  selfbound.}
\label{fig:spectral_mr}
\end{figure}

The determination of the mass-radius relation of a neutron stars are model
dependent so far. The outermost layer of the neutron star, the
atmosphere, needs to be modeled in order to fit to the observed x-ray
spectra. Presently, the atmosphere of isolated neutron stars is not
understood, as the optical flux is not compatible with an
extrapolation of the observed x-ray spectra. The extracted radius is
the one as observed by an observer at infinite distance to the neutron
star and is defined by the true radius and the mass of the star as
\begin{equation}
R_\infty = \frac{R}{\sqrt{1-R_s/R}}
\end{equation}
with the Schwarzschild radius $R_s=2GM$.  The most prominent example
is the isolated neutron star RXJ1856.5-3754 for which a constraint on
the radiation radius of $R_\infty > 17$~km~$(d/140 pc)$ was
inferred \cite{Trumper:2003we,Ho:2007gs}. For the given redshift of
$z\approx 0.22$ this radiation radius results in $M\approx 1.55
M_\odot$ and $R=14$~km.  The largest error resides in the distance $d$
of the pulsar to Earth, so that in principle any mass between
$1.1M_\odot$ and $2M_\odot$ with radii of 10 km and 18 km,
respectively, seems possible.

\begin{figure}
\centerline{
\includegraphics[height=0.4\textheight]{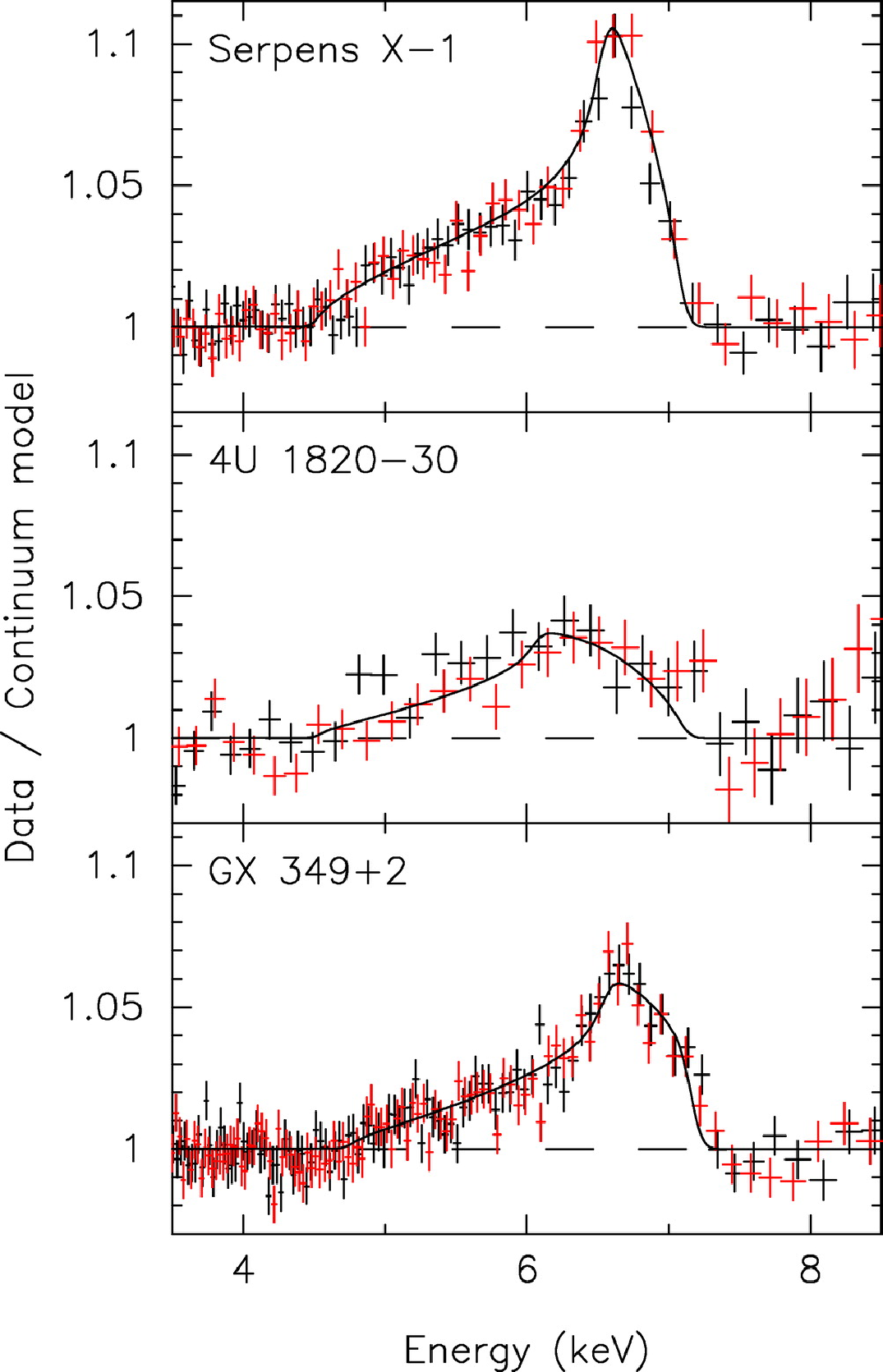}\qquad
\includegraphics[height=0.4\textheight]{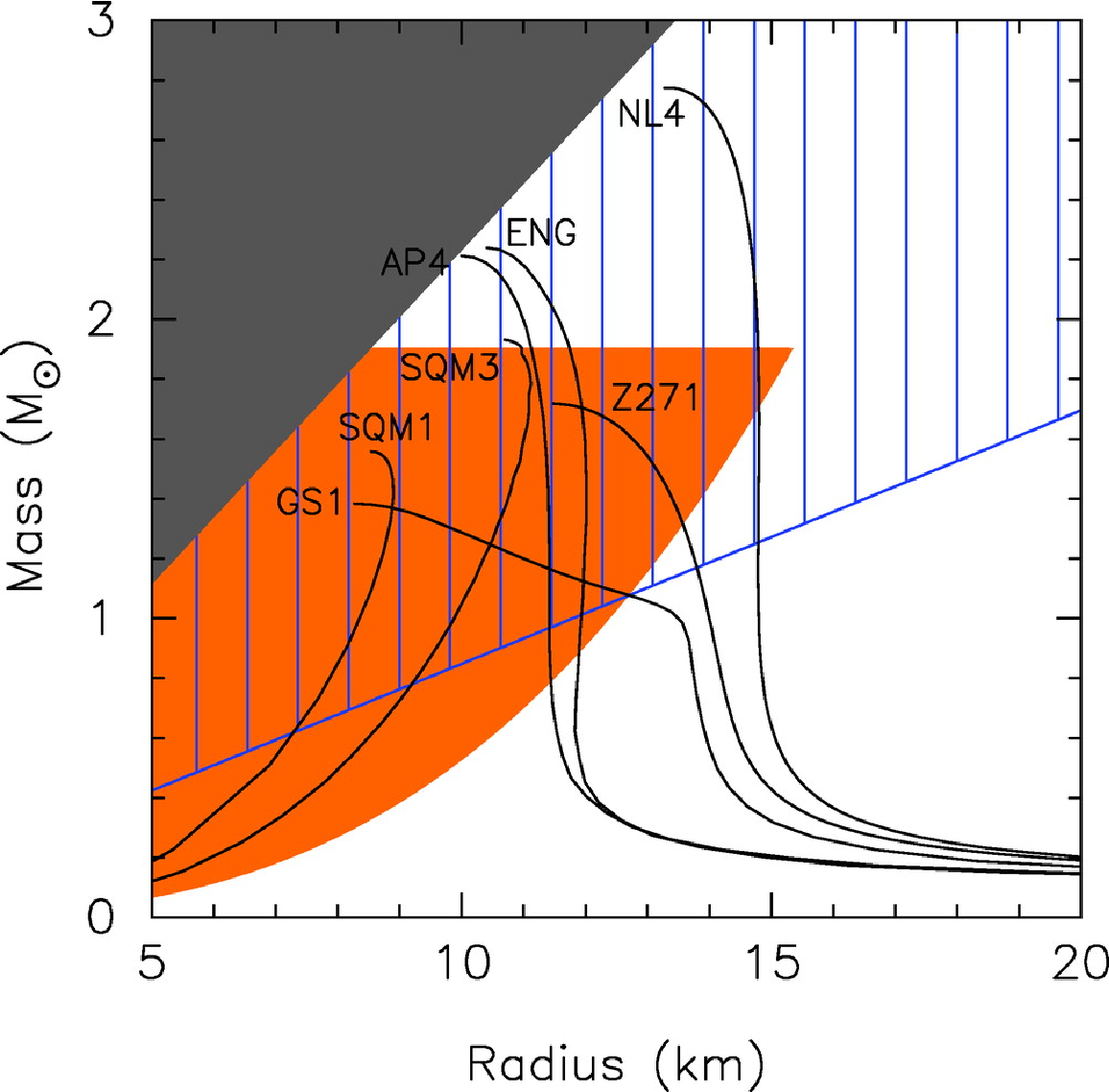}}
\caption{Left: Redshifted iron lines from several low-mass x-ray
  binaries. Right: Constraint on the mass-radius relation for neutron
  stars from the maximum redshift measured for the iron lines,
  indicated by the striped region. The pie-shaded filled area is the
  region in mass and radius compatible with the measurements of
  quasi-periodic oscillations (QPOs) of accreting neutron stars
  (figures are taken from \cite{Cackett:2007}).}
\label{fig:ironlines}
\end{figure}

Accreting neutron stars can be sources of x-ray bursts as matter is
heated up during accretion and falls onto the neutron star. For the
x-ray burster EXO 0748-676 redshifted spectral lines have been
extracted in the aftermath of an x-ray burst \cite{Cottam:2002cu}
seemingly originating from the surface of the compact star. Follow-up
observations of another burst in 2003 could not confirm this finding
\cite{Cottam:2007cd}. A model analysis of the x-ray burst led to
rather tight constraints for the mass and radius of the compact star
of $M\geq (2.10\pm 0.28)M_\odot$ and $R\geq (13.8\pm 1.8)$~km
\cite{Ozel:2006km} which are based on the redshift measurement of
\cite{Cottam:2002cu}.  However, a detailed multiwavelength analysis
concluded that the mass of the compact star is more compatible with
$1.35M_\odot$ than with $2.1M_\odot$ \cite{Pearson:2006zy}. In any
case, compact star masses and radii constraints as given by
\cite{Ozel:2006km} still allow for the possible existence of quark
matter in the core of the compact star \cite{Alford:2006vz} contrary
to the strong claims made in \cite{Ozel:2006km}. Spectral fits to
accreting neutron stars in quiescence were hampered by the fact that
the surface gravity of the neutron stars was not adjusted consistently
but was fixed. Starting with ref.~\cite{Rybicki:2005id} consistent
spectral fits became available, first applied to the neutron star X7
in the globular cluster 47~Tuc. The constraint on the mass and radius
of the neutron star are quite loose.  The authors quote strong limits
for the mass by fixing the radius and vice versa which should be taken
with care. A glance at their exclusion plot shows that for a radius of
$R\approx 14$~km any mass between $0.5M_\odot$ and $2.3M_\odot$ is
allowed by the fit. Spectral modeling of the neutron stars in
$\omega$~Cen and M13 can be found in \cite{Webb:2007tc} with similarly
large errors. Note, that the combination of several spectral modeling
results on the constraint for the mass-radius relation of neutron
stars that the mass-radius curve has to pass through all the different
regions allowed by the spectral fits somewhere and not that the curve
has to strike the common area of all fits. The spectral fit of the
neutron star in M13 demands that the radius should be smaller than
about 12~km with a mass smaller than $1.8M_\odot$ on the 99\%
confidence level (see fig.~\ref{fig:spectral_mr}).

Interestingly, there are constraints on the radius of the neutron star
which mostly rely on General Relativity. A broadened iron-line
characteristic for redshifted iron-lines from an accretion disk have
been measured for three low-mass x-ray binaries \cite{Cackett:2007}.
The lower endpoint in the energy of the broadened line determines the
maximum redshift and therefore the inner radius of the accretion disk
relative to the Schwarzschild radius of the neutron star. It turned
out that neutron stars should have a radius smaller than $R=(7-8)GM$
which for an assumed mass of $1.4M_\odot$ results in $R<14.5-16.5$~km
(see Fig.~\ref{fig:ironlines}).  Limits on the compactness, the
mass-to-radius ($M/R$) ratio, can be extracted rather model
independent from the profile of thermal x-rays emitted from hot spots
on the surface of weakly magnetized millisecond pulsars as
demonstrated in \cite{Bogdanov:2006zd,Bogdanov:2008qm}. For the pulsar
J0030+0451 the neutron star must have $R/R_s>2.3$ which for
$M=1.4M_\odot$ demands for a radius of $R>9.5$~km.

A certain class of binary neutron stars emits quasi-periodic
oscillations and are dubbed therefore QPOs. The neutron star is
accreting material from a companion star which causes these
oscillations. General Relativity predicts that there exists an
innermost stable circular orbit (ISCO) which for a spherical system is
given by $R_{\rm isco} = 3R_s = 6GM$ in the Schwarzschild-metric when
effects from the rotation of the neutron star can be neglected. If an
innermost stable circular orbit of these accreting systems can be
detected the radius of the neutron star has to be lower than this
limit providing a rather stringent constraint on the properties of
neutron stars. The highest QPO frequency measured so far is for the
system 4U 0614+091 with a frequency of 1330 Hz
\cite{vanStraaten:2000}. A stable circular orbit within the
Schwarzschild metric is given by $\Omega^2 = M/r^3$.  The radius of
the neutron star has to be lower than that giving the constraint
\begin{equation}
R \leq R_{\rm orb} = \left( \frac{GM}{4\pi^2 \nu_{\rm QPO}^2} \right)
\end{equation}
where $\nu_{\rm QPO}$ is the highest measured QPO frequency.
General Relativity demands that the stable circular orbit must be larger than
the ISCO, $R_{\rm orb} \leq R_{\rm isco}$.  The combination of the two
conditions cuts out a wedge-like shape in the mass-radius diagram
which is bounded by a maximum mass and a corresponding radius. The
mass-radius curve for neutron stars has to pass this region
\cite{Miller:1998}, see also the mass-radius plot in
Fig.~\ref{fig:ironlines}. There have been several claims that indeed
the innermost stable circular orbit has been detected, most notably
and recently for the QPOs measured for 4U 1636
\cite{Barret:2007df} arriving at a neutron star mass of $2.0M_\odot$.
A phase resolved spectroscopy with the VLT recovers a smaller neutron
star mass range of 1.6 to 1.9$M_\odot$ prevailing any firm conclusion
about a massive compact star.

There are rather model independent constraints on the maximum mass
possible for neutron stars which rely just on general relativity and
causality for the nuclear equation of state. Let us assume that the
nuclear equation of state is known up to some fiducial energy density,
be it from the measurement of the properties of nuclei or from the
determination of the in-medium properties of hadronic matter as
generated in heavy-ion collisions. Then the high-density equation of
state above that fiducial energy density can be limited by demanding
that the hydrodynamical speed of sound can not exceed the speed of
light. For the simplest case this criterion reads for the equation of
state that
\begin{equation}
p = c_s^2 \epsilon \quad \mbox{with} \quad c_s^2\leq c^2 \quad .
\end{equation}
Even if the nuclear equation of state is unknown above some energy
density, the pressure can not rise more rapidly than given by the
limiting case $p=\epsilon$, i.e.\ the nuclear equation of state can
not be stiffer than that due to causality. In turn, the stiffest
possible equation of state provides the maximum possible mass
configurations for compact stars, as it gives for a given energy
density the maximum possible pressure which can counterbalance the
pull of gravity. Now one can assume that one knows the nuclear EoS up
to some fiducial energy density. For higher energy densities one
adopts the stiffest possible equation of state. The resulting maximum
mass is the highest neutron star mass allowed by causality and by the
nuclear EoS fixed up to a certain energy density.  There exists
scaling relations for the TOV equations.  It can be shown that the
maximum possible mass scales in the following way with the fiducial
energy density $\epsilon_f$ where one switches from the nuclear EoS to
the stiffest possible EoS:
\begin{equation}
M_{\rm max} = 4.2 M_\odot \left(
\frac{2.5\cdot 10^{14}\mbox{ g cm}^{-3}}{\epsilon_f}
\right)^{1/2}
\end{equation}
where $\epsilon_0= m_N\cdot n_0 = 2.5 \cdot 10^{14}$g cm$^{-3}$
corresponds to a nuclear (number) density of $n_0=0.16$~fm$^{-3}$. The
prefactor of $4.2M_\odot$ is determined numerically and depends on the
nuclear EoS adopted. There exists several investigations on this
maximum mass constraint using a different nuclear EoS which come to
about similar numerical values for the maximum mass
\cite{Rhoades:1974fn,Hartle78,Kalogera:1996,Akmal:1998cf}. Note that
the nuclear EoS utilized in these works are constrained by
nucleon-nucleon interactions and the properties of nuclei, so the
nuclear EoS is probed only up to normal nuclear matter density.
Usually, a higher fiducial energy density is given, about twice the
normal value, which is unjustified and would result in a maximum mass
constraint of around $3M_\odot$, which is actually the standard value
quoted in the literature. We argue that the nuclear EoS can not be
reliably fixed from the properties of the nucleon-nucleon interaction
and of nuclei above normal nuclear density at present so that the
correct mass limit from nuclear models is given by about $4.2
M_\odot$. The determination of the nuclear EoS at supranuclear
densities as done with subthreshold kaon production gives tighter
constrains on the maximum mass, as the fiducial density can be
increased to about 2 to 2.5 times normal nuclear matter density,
so that the maximum possible mass for a neutron star would be limited
to about $2.7-3.0 M_\odot$.

In addition, there is another ingredient to the nuclear EoS for
neutron stars which is difficult to determine at present: the density
dependence of the nuclear asymmetry energy (for a recent review on the
role of the asymmetry energy for nuclei and neutron stars see
\cite{Steiner:2004fi}). It is noteworthy, that there seems to be a
general trend for the mass-radius relation of compact stars for modern
realistic nuclear equation of states. The lower end of the neutron
star branch is determined by the rather well known low-density
equation of state and is located around a mass of 0.1~$M_\odot$ and a
radius of about 250~km (see e.g.\ \cite{BPS,Macher05}). The maximum
density is just above the critical density for the onset of
homogeneous neutron star matter, about half normal nuclear matter
density. As the neutron star material consists of a lattice of nuclei
immersed in a gas of electrons and (free but interacting) neutrons
below that density, the neutron star is mainly solid not liquid. The
low-density equations of state are usually taken from
\cite{BPS,Negele73}, see \cite{Ruster:2005fm} for an update of the so
called BPS equation of state \cite{BPS}. At one to three times normal
nuclear matter density, interactions start to give a sizable
contribution to the pressure. The resulting equation of state can be
approximated by a polytrope of the form $p \sim n^2 \sim \epsilon^2$.
The exact critical density where the overturn happens is controlled by
the strength and the density dependence of the asymmetry energy. The
mass-radius relation for a polytrope of the above form is well known
and quite simple: the radius becomes independent of the central
density and therefore of the mass of the neutron star. Indeed, one
finds in more sophisticated approaches, that the mass increases
drastically between a rather narrow window in radius of typically 10
to 15~km, see the reviews \cite{Lattimer:2000nx,Lattimer:2006xb}.
Common nonrelativistic approaches to the neutron star matter equation
of state have lower radii than relativistic approaches, as the density
dependence of the asymmetry energy in relativistic models is
generically much stronger than in the nonrelativistic ones. The
observation of the relation between the radius of a neutron star and
the asymmetry energy led to the idea to determine the asymmetry energy
by measuring the neutron radius of led to learn something about the
mass-radius relation of neutron stars \cite{Horowitz:2000xj}. However,
we point out that the central density for neutron star masses observed
so far are well above normal nuclear matter density. One needs
actually to know the asymmetry energy at about three times normal
nuclear matter density. This density regime could be reached by
heavy-ion collisions at a few GeV bombarding energy, where particle
ratios as the subthreshold $K^-/K^+$ ratio \cite{Ferini:2006je} or the
$\pi^-/\pi^+$ ratio measured with the FOPI spectrometer
\cite{Reisdorf:2006ie,Xiao:2008vm} could serve as a probe of the
isospin dependent forces at high densities.

At high densities the nuclear equation of state is not only
essentially unknown but also the overall composition and structure
could be totally different. New exotic particles and phases can appear
which alter not only the global properties of compact stars, the total
mass and radius, but also the cooling evolution and the stability
against the emission of gravitational waves or the delayed collapse to
a black hole.

Many different model approaches predict that hyperons are present in
neutron star matter around twice normal nuclear matter density. The
composition of hyperons depends crucially on the hyperon-nucleon
interactions, in particular on the hyperon self-energies. Hyperons can
be abundantly present so that a neutron star is in this case more
aptly dubbed a 'giant hypernucleus' \cite{Glendenning:1984jr}. The
impacts of hyperons on the properties is manyfold. Most importantly,
the maximum possible mass is drastically reduced, even when many-body
effects and a repulsive interactions between hyperons are taken into
account \cite{Glendenning:1991es}. If there is a phase transition to
hyperon-rich matter in the core of neutron stars, a new stable branch
in the mass-radius curve can be present which are compact stars with
similar radii but smaller radii \cite{SchaffnerBielich:2002ki}. But
also the cooling is affected as hyperons open new additional cooling
processes which are controlled by the hyperon-hyperon interactions.
Hypernuclear experiments measuring double hypernuclei at J-PARC and at
the PANDA experiment at FAIR, GSI Darmstadt will shed more light on the
strength of the hyperon-hyperon interactions in the near future.
Bound states of hyperons could be formed in relativistic heavy-ion
collisions as dozens of hyperons are produced in a single event (see
e.g.\ \cite{SchaffnerBielich:1999sy}). The study of those systems will
be crucial in determining the hyperon-hyperon interaction and the
hyperon composition of neutron stars in the interior. For a review on
the relation between hypernuclear physics and neutron stars see
\cite{SchaffnerBielich:2008kb}.

\begin{figure}
\centering
\includegraphics[width=0.9\textwidth]{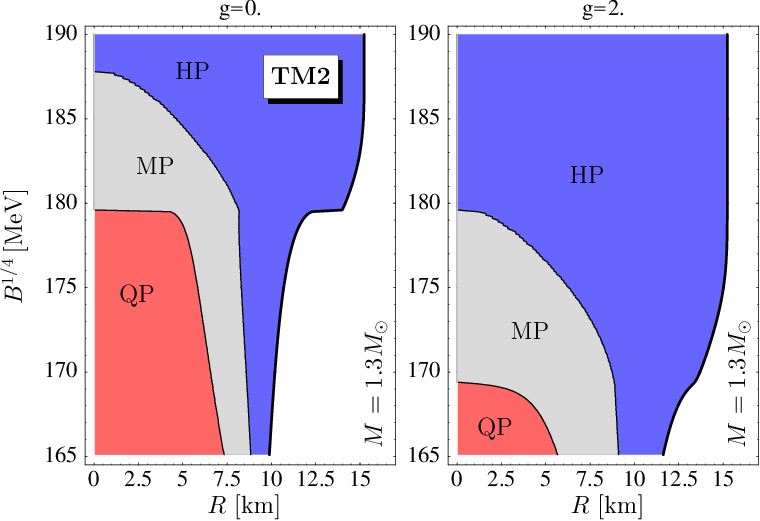}
\caption{Cut through a neutron star with quark matter for different
  values of the MIT bag constant without (left plot) and with
  corrections from a hard dense loop calculation (right plot). The
  figure is taken from \cite{Schertler:2000xq}. QP: Quark phase, MP:
  Mixed phase, HP: Hadronic phase.}
\label{fig:hybridstar}
\end{figure}

Around three to four times normal nuclear matter density another sort
of strange hadron could be present in the dense interior of compact
stars: Antikaons which form a Bose condensate. Antikaons are formed in
beta-stable matter by transforming electrons to negatively charged
antikaons, $e^- \to K^- + \nu_e$, where the neutrino is emitted and
acts as an additional cooling agent \cite{Brown92}. The loss of
electron pressure and the nonexisting pressure contribution from the
Bose condensate destabilizes the compact star matter so that the
maximum mass of neutron stars is drastically reduced giving rise to
stellar mass black holes, the Bethe-Brown scenario \cite{BB94}.
Crucial input to the scenario of antikaon condensation is the highly
attractive potential felt by antikaons in dense matter, a scenario
which can be probed by subthreshold production of antikaons in
relativistic heavy-ion collisions as measured by the KaoS
collaboration \cite{Kaos97,Sturm:2000dm,Forster:2007qk}. For the
imminent correlation between antikaon condensation and antikaon production
in heavy-ion collisions see \cite{Li97L,Li97}. The in-medium potential
of antikaons can not be reliably extracted from the production rates,
as matter produced in heavy-ion collisions has different properties
than neutron star matter as (see e.g.\ \cite{SKE00} for a detailed
investigation on this point). The strong in-medium potentials for
antikaons also enhances the in-medium cross section so that the
production rates saturate. The correlated emission of antikaons might
serve as a better observable to probe the in-medium potential of
antikaons. We point out again that the subthreshold production of
kaons ($K^+$) is an excellent tool for constraining the nuclear equation
of state above normal nuclear matter saturation density as discussed
above in more detail. However, in contrast to antikaons the $K^+$ feels a
repulsive potential in dense matter so that kaons are unlikely to be
present in neutron star matter.

Finally, the extreme densities in the core of neutron stars can reach
the point in the QCD phase diagram where matter is converted from the
hadronic chirally broken phase to the chirally restored quark matter
phase (here for simplicity we denote the new phase as being quark).
The phase transition to quark matter in compact star matter can have
profound consequences for the physics of neutron stars and
core-collapse supernovae. The physics of strange quark matter and
quark stars have been reviewed in \cite{Weber:2004kj}. Many signals
for the presence of quark matter and the QCD phase transition have
been proposed in the literature also for explosive processes in
astrophysics as for gamma-ray bursts, gravitational wave emission from
neutron star mergers (for a recent review we refer to
\cite{SchaffnerBielich:2007mr}). The amount of quark matter in the
core of compact stars is still an open question and best illustrated
in Fig.~\ref{fig:hybridstar}.

The astrophysical constraints on the nuclear equation of state and the
quark matter equation of state have been investigated in more detail
in \cite{Klahn:2006ir,Klahn:2006iw} including also observables from
heavy-ion experiments. Some of the astrophysical input data has been
revised in the mean-time and some can not be taken as serious
constraints as discussed above. If limits from astrophysical data are
taken firmly and as proposed in the literature, in particular if one
assumes a maximum mass of a neutron star of $2M_\odot$ or more, then
it turns out that none of the purely hadronic equation of state can
fulfill all the constraints imposed. This finding points to two
important statements: astrophysical data can indeed give strong
constraints on the nuclear equation of state but it has to be taken
with great caution. In the near future with the advent of new
detectors one will get a much tighter grip on the properties of the
nuclear equation of state at high densities from astrophysical
observations.


\end{document}